\documentclass[a4paper,12pt]{article}

\usepackage{graphicx}
\usepackage{wasysym}

\begin{document}

\begin{titlepage}
\title{Molecular switch controlled by pulsed bias voltages}

\author{Velimir Meded, Alexei Bagrets, Andreas Arnold, and Ferdinand Evers
\thanks{PD Dr. Ferdinand Evers\newline
Institut f\"ur Nanotechnologie\newline
Forschungszentrum Karlsruhe\newline
D-76021 Karlsruhe, Germany\newline
Institut f{\"u}r Theorie der Kondensierten Materie\newline
 Universit{\"a}t Karlsruhe\newline
D-76128 Karlsruhe, Germany\newline
E-mail: Ferdinand.Evers@int.fzk.de\newline
Dr. Velimir Meded, Dr. Alexei Bagrets \newline
Institut f\"ur Nanotechnologie\newline
Forschungszentrum Karlsruhe\newline
D-76021 Karlsruhe, Germany\newline
Dr. Andreas Arnold\newline 
Institut f{\"u}r Theorie der Kondensierten Materie\newline
Universit{\"a}t Karlsruhe\newline
D-76128 Karlsruhe, Germany}}


\maketitle
\clearpage
 
\begin{abstract}
It was observed in recent experiments that the current-voltage
characteristics (IV) of BPDN-DT (bipyridyl-dinitro 
oligophenylene-ethynylene dithiol) can be switched 
in a very controlled manner between ``on'' and ``off''
traces by applying a pulse in a bias voltage, $V_\mathrm{bias}$.
We have calculated the polaron formation energies to check
a frequently held belief,
namely, that the polaron formation can explain the observed 
bistability. 
Our results are not consistent with such a mechanism.
Instead, we propose a conformational reorientation.
The molecule carries an intrinsic dipole moment which couples to $V_\mathrm{bias}$. 
Ramping $V_\mathrm{bias}$ exerts a force on the dipole
that can reorient (``rotate'') the molecule from the ground
state (``off'') into a metastable configuration (``on'') and back.
By elaborated electronic structure calculations, we identify
a specific path for this rotation through the molecule's
conformational phase space. We show that this path has 
sufficiently high barriers to inhibit
thermal instability but still the molecule can be switched in 
the voltage range of the junction stability. The theoretical IVs 
reproduce qualitatively the key experimental observations. 
We propose, how the alternative mechanism of conductance switching 
can be experimentally verified.
\end{abstract}

\end{titlepage}

\section{Introduction}

During the past few years, experiments in the field of 
``Molecular Electronics'' have experienced significant improvements.
This made possible the observation of Coulomb blockade \cite{coulomb} and 
Kondo effects \cite{kondo} in molecular transistor geometry. 
Improved statistical methods to access a conductance of molecular 
junctions have been developed \cite{statist,chemist}.
A spectacular manifestation of molecular low energy excitations
has been detected \cite{excitations} in inelastic current spectroscopy.

Aiming for potential applications, Molecular Electronics
holds the vision that functional devices like memory elements, 
switches, transistors etc. may be realizable by designing 
suitable molecular complexes. In this spirit, a considerable 
effort to design and test molecular systems 
with a controlled switching behavior has been made \cite{tao06,weibel07}.
The challenge here is a ubiquitous one in the field of Nanotechnology,
namely how to build and manipulate a device or a material with full
control down to the atomic level. Therefore, one may expect that a
solution in one subfield will cross-fertilize exciting developments in many
others. 

The proposed prototypes for single molecule memory elements 
can be sorted according to  
which physical degree of freedom underlies the switching bistability. 
The three principle categories are:
(I) switches with different (meta-)stable charge configurations 
    \cite{haiss03, chen04, xu05, li07, galperin05,ryndyk08};
(II) conformational switches with molecules exhibiting different 
     stable isomers \cite{kornilovitch02, luo02, liang07, whalley07}; 
(III) conformational switches with molecular reorientations against the contact 
     atoms \cite{lewis05}. 
Once it is known how to operate a single molecule, arrays of molecules
can be addressed as well, which eventually may form an entirely new
class of molecular hybrid-materials. 

It is very encouraging, that in recent experiments switching 
has been successfully demonstrated by   
Blum {\it et al.}\ \cite{blum05}, 
Keane {\it et al.}\ \cite{keane06} 
and L\"ortscher and Riel \cite{loertscher06}. 
They have been using a molecular wire of the "Tour-type", BPDN-DT, 
see Fig.~\ref{f1}, which was investigated in several 
experimental and theoretical works, before \cite{lewis05,elbing05,he06}.
This particular set of experiments has two intriguing features:
first, switching is established in a two-terminal device, just employing a
pulse in the bias voltage, and second the effect is very stable, so that
it has even been observed at room temperature. 

Specifically, the BPDN-DT wire shows the following characteristic behavior in 
experiment \cite{loertscher06}: 
with a very slow (adiabatic) change of the bias,
$V_{\rm bias}$, the current voltage characteristic (IV) 
is fully reversible for up-sweeping and down-sweeping. 
However, with faster sweeping rate the adiabatic regime is left and 
a hysteretic behavior is seen in the IV if $V_{\rm bias}$ is 
tuned beyond $\approx 1$V. The new IV-curve signalizes the existence 
of a second (meta-)stable state of the molecular junction. 
One returns to the original curve only by repeating 
the procedure with inverted bias. 

The physical origin of bistability in these experiments 
has not yet been identified up to now. He {\it et al.} \cite{he06} 
attribute it for their experiment in electrochemical environment 
to a change in the oxidation state of the molecule. 
A polaron caused bistability was put forward
by Galperin {\it et al.} \cite{galperin08} also as a possible explanation 
of switching observed in the break junction experiment
by L{\"o}rtscher {\it et al.} \cite{loertscher06} who
investigated a freely suspended molecular bridge.
On the other hand, Keane {\it et al.} \cite{keane06} argue that the
formation of a polaron is not a likely occurrence in their electromigrated 
break junction experiment. They suspect that rather
a bias driven modification of the contacts may be
responsible.  

Identifying a possible switching mechanism operating for 
Tour-type mo\-le\-cules is the aim of this work. 
Our calculations suggest the following picture: the BPDN-DT molecule in vacuum
or in between two contacts, Fig.~\ref{f2},  
is not susceptible to charging. Unless it 
is stabilized by counter-charges that are located very
close to the bipyridine unit, any excess charge recombines 
with its images on the electrodes (Fig.~\ref{f3}). 
This makes polaron-formation (class I) unlikely in the experiments 
\cite{blum05,keane06,loertscher06}
in support of an earlier claim \cite{keane06}. 

However as we shall demonstrate, the key property for switching,
bistability, can originate from a rotational degree of freedom associated with 
BPDN-DT molecule contacted to electrodes. 
The inertia that is driving the rotation results from the action of 
the electrical field (associated with $V_{\rm bias}$) 
on the dipole moment of the dressed bipyridine unit. 
Two stable configurations can be reached by a double axis
$\pi$-rotation, see Fig.~\ref{f4} (classes II \& III). 
Our calculations suggest, that the rotation can be performed in such a
way, that the rotational barrier is high enough --- so both states are
stable against temperature --- but at the same time 
this barrier is low enough so that 
one state rotates into the other under bias voltages $\sim 1$V (Fig.~\ref{f5}). 
The theoretical IV curves (Fig.~\ref{f6}), which one
obtains for either state, closely resemble the 
experimental findings \cite{blum05,keane06,loertscher06}. 

\section{Model and Method} 
The model setup used for the DFT
transport calculations is depicted in Fig.~\ref{f2}. 
The Au$_{14}$-clusters mimic the electrodes. Computations based 
on the density functional theory (DFT)
have been performed  with the real-space based 
package {\small TURBOMOLE} \cite{turbomole}. 
Optimized basis sets of triple-$\zeta$ quality 
including polarization functions have been used
(exception: IV with a split-valence basis set of double-$\zeta$ quality)
\cite{schaefer92,def2-TZVP}. The exchange correlation (XC) functional  
BP86 \cite{BP86a,BP86b} was employed. Charging analysis and 
relaxations were checked against a hybrid functional, B3LYP \cite{B3LYP}.  

\section{Absence of polarons} 
In order to investigate the possibility of polaron formation, 
we have considered the core region of the molecule, only, replacing
the Au$_{14}$ cluster by an H-atom. We compared
the electronic structure for the uncharged ($Q{=}0$) and the 
charged species ($Q{=}\pm 1$ in units of the electron charge $|e|$). 
Energies for the highest occupied (lowest unoccupied) molecular level, 
HOMO (LUMO), are given in Fig.~\ref{f3}. We first discuss the
molecule in the gas phase, ``free'' molecule. The data indicate, that
the LUMO of the positively charged ion ($Q{=}+1$) has an energy
$E_{\rm LUMO}(Q{=}+1)\approx -8.1$eV. Upon approaching a metal 
surface, $E_{\rm LUMO}(Q{=}+1)$ will increase due to the interaction
with the image charge. Assuming, that the excess
charge distribution is arranged about the center of the molecule with
an extension much smaller than the molecule length, 
one can give an estimate of the energy change 
by employing an image charge analysis.
We obtained an energy $E_{\rm LUMO}(Q{=}+1)\approx -7.2$eV for the
LUMO after screening. Since it is certain \cite{umeno07}, 
that the workfunction of any uncharged Au-surface is 
above $-6$eV and below $-5$eV, one can conclude that the positively
charged ion will be neutralized as soon as it makes contact to a
Au-surface. By a completely analogous argument for the HOMO of the
anion (see Fig.~\ref{f3}), we deduce that also the 
molecule with $Q{=}-1$ is unstable near a Au-surface.

All calculations have been performed with a fully relaxed molecular
structure; the charged molecule undergoes a conformational
change, which mostly affects the NO$_2$-groups. However, our
calculations do not give any indication, that a geometrical
deformation could stabilize the ion in the vicinity of the electrodes, 
see Fig.~\ref{f3}. More precisely, the energy change of the anion's
HOMO (cation's LUMO) upon structure relaxation is not sufficient in order to
induce a level crossing with the Fermi energy. 
This implies, that the contacted BPDN-molecule 
does not exhibit a polaron inside a trivial vacuum bounded by the two
electrode plates. This is the case in the break junction
experiments \cite{keane06,loertscher06}. Our conclusion does 
not necessarily apply to experiments in an electrochemical environment
\cite{he06} if this has a sufficiently large dielectric response. 


\section{Rotations and bistability}
After we have ruled out charge degrees of freedom as a likely 
origin for bistability, we now turn our attention to mechanisms related
to charge neutral conformational changes. Since the control molecule, 
Fig.~\ref{f1}, did not show hysteresis 
\cite{blum05,loertscher06}, bistability should involve a (charge neutral)
conformational modification of the BPDN-functional unit. 
An important second requirement is that an external electric 
field (realized via $V_{\rm bias}$) must be able to address the
putative degree of freedom. Since the monopole moment of the
functional unit remains zero, it should be the force on its dipole 
moment, $p_0$, generated by the NO$_2$-groups that pushes one 
configuration into the other and back; according to our calculations 
(employing BP functional) ${p_0}\simeq$~3.0 Debye. Unlike the BPDN-DT, 
the reference molecule (BP-DT) enjoys the inversion symmetry and, 
therefore, has a vanishing intrinsic dipole moment. We take 
this an explanation why its IV does not exhibit hysteretic behavior. 
 
By this reasoning, one is led to look for rotations that take either 
the entire molecule or at least its BPDN-unit over from a stable
ground state configuration (GC, ``off'') to another (meta-)stable 
one (MC, ``on''). Any such rotation corresponds to a path in the atomic 
configuration space of the (extended) molecule. Apart from
bistability, the optimal path has to satisfy a number of constraints
in order to be a viable candidate for realizing a molecular switch. 
(i) The energy barrier between GC and MC, $\Delta E^*$, should exceed
temperature, $\Delta E^* \gg T$, to avoid uncontrolled thermal switching. 
(ii) In external fields the optimal path should have a continuous 
deformation so that GC at forward bias, say, and MC at reverse bias 
become unstable in the sense that $\Delta E^* \apprle T$. 
The instability should be reachable at ``switching" voltages, 
$V_{\rm on}$, still tolerable for the molecular junction, in 
experiments $V_{\rm on}\sim 1$~Volt.
(iii) GC and MC should have two IV curves, $I_{\rm on}$ and $I_{\rm off}$, 
which can be discriminated from each other. 

Giving typical switching voltages $V_\mathrm{on}\sim 1$~Volt, 
one can estimate the energy barrier $\Delta E^*$ separating two states 
at zero bias. Assuming that the molecule (length $d{\approx}25$~\AA) 
forms an angle ca. 45$^\circ$ with the bias electric field, we have 
${\cal E}_{\rm on}{\simeq} \sqrt{2} V_{\rm on}/d$. 
Once a dipole, $p_0$, is flipped in the bias field, 
the system's energy gain is $\Delta E_\mathrm{dipole} {\simeq} 2 p_0 
{\cal E}_{\rm on} /\sqrt{2}$, which should be of order $\Delta E^{*}$ 
if energy mismatch between GC and MC is small compared to $\Delta E^{*}$.
Using $p_0 \simeq 3$~Debye, one arrives to an estimate $\Delta E^{*} \sim 
\Delta E_\mathrm{dipole} {\approx}50$~meV.

After an extensive search, we have found a family of paths which is 
best described as a consecutive one parameter (angle $\theta$) rotation 
about two axes, see Fig.~\ref{f4}. The family parameter is the
(average) angle, $\Phi$, that is formed between the longitudinal
axis of the molecular wire and the (hypothetical) surface normal of
the electrode. 

Consider a variation of ground state energy of the system, 
$\Delta E_\mathrm{tot}(\theta)$, with .rotation angle $\theta$.
As may be inferred from Fig.~\ref{f5} (trace 0V), 
the example path with $\Phi{\approx}70^{o}$ satisfies 
condition (i), since the barrier between GC ($\theta{=}0$) 
and MC ($\theta{\approx}\pi$) is 
$\Delta E^*{\approx}\Delta E_{\rm tot}(\pi/2){\sim}100{\rm meV}{\approx}1160$K 
which is in rough agreement with the above estimate. 
(Notice, that the double axis rotation is crucial to obtain consistent 
energies. Namely, unbalanced single rotations, {\it e.g.}\ about 
the S-S axis or the C-C axis near the functional unit, give much higher 
barriers, $\sim$ 200~meV, and thus are ruled out.)

Furthermore, the evolution of the traces depicted in Fig.~\ref{f5} 
under the applied homogeneous electric field, $\cal{E}_{\rm h}$,
shows that for a path with $\Phi{\approx}70^{o}$ 
also condition (ii) is met: the value of $\cal{E}_{\rm h}$ 
at which either GC or MC becomes unstable corresponds to an effective voltage
$|V_{\rm on}|{\approx}2 \div 4$~Volt. 

Finally, we can show that also (iii) is
fulfilled. To this end we have determined the IV curve for the
situations GC and MC, again at $\Phi{\approx}70^{\rm o}$, with our 
home made transport package employing the non-equilibrium Green's
function method \cite{evers_arnold}.

\section{IV-characteristics}  
\begin{figure}[t]
\begin{center}
\includegraphics[width=0.75\textwidth]{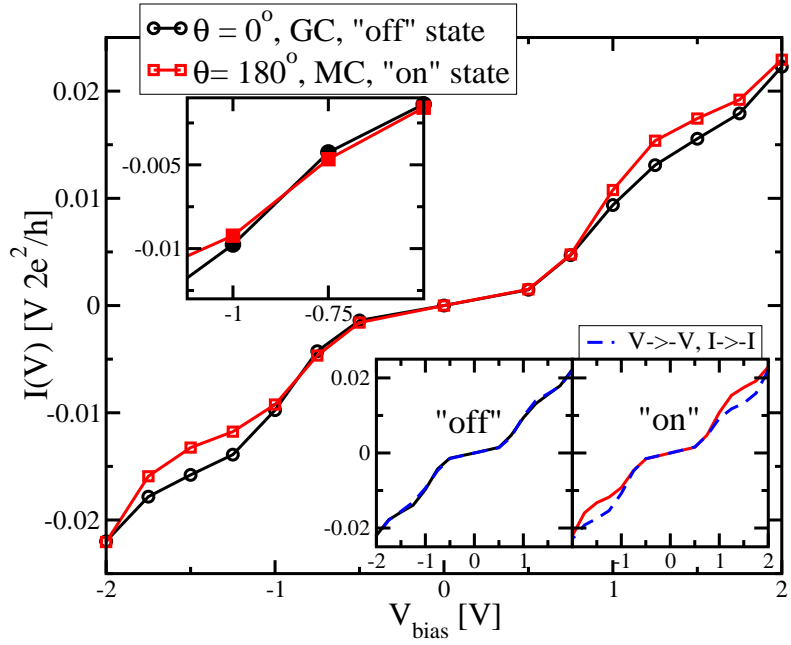}
\end{center}
\caption{Current voltage trace for the BPDN-DT
     molecule at $\Phi{\approx}70^{\rm o}$ for the ground state
     configuration, GC ($\circ$),
     as well as for the meta-stable one, MC ($\Box$).
     Inset upper left: blow up of main plot near small negative
     $V_{\rm bias}$.
     Inset lower right: Behavior of IV curves under the transformation
     $V_{\rm bias}\to -V_{\rm bias}, I\to-I$, dashed line
     (left ``on'', right ``off'').
}
\label{f6}
\end{figure}

The theoretical IV-curves, Fig.~\ref{f6}, clearly
displays two different IV traces which allow to 
read out the state of the molecular bit, ``on'' or ``off''. 
They exhibit key features, which now we discuss:

(a) A step-like increase in the current is observed (near 1V). It  
is mainly due to the molecular orbital HOMO$^*$ entering the voltage window. 

(b) Under the transformation 
\begin{equation}
{\cal I:} \quad V_{\rm bias}\to-V_{\rm bias}, I\to-I
\label{e1}
\end{equation}
both IVs are nearly invariant, see insets Fig.~\ref{f6}; 
Eq. (\ref{e1}) implies that $dI/dV$ is an even function 
of $V_{\rm bias}$. Generally speaking, 
${\cal I}-$invariance is an exact property of
non-interacting (i.e.~non-polarizable) electron 
systems.\footnote{In our discussion we ignore any structure 
in the density of states, $\varrho(E)$, 
of the electrodes; i.~e. we assume 
that $\varrho(E)$ is independent of energy.}\ 
In interacting systems, charge localized on the molecule is redistributed as a 
response to $V_{\rm bias}$. The polarization produces a change in the effective
potential that feeds back into the current carrying orbitals 
thus giving rise to second order effects, $V_{\rm bias}^2$, on the current. 
Details of the redistribution depend strongly on contacts and 
the orientation of the molecule in the bias induced $E$-field
(in particular its sign). In the present case the molecular junction
has an approximate inversion symmetry (see Fig.~\ref{f4}) and
therefore violation of ${\cal I}$ is weak.

(c) It is important for understanding the theoretical IV 
to notice, that symmetry violation is stronger with ``on-''
than with the ``off-''trace (inset Fig.~\ref{f6}), so that 
$I_{\rm on}>I_{\rm off}$ at positive $V_{\rm bias}$ but 
$|I_{\rm on}|<|I_{\rm off}|$ at sufficiently large reverse 
biases $|V_{\rm bias}| > V_{\rm c} $ 
(dashed trace in Fig.~\ref{f6} is always above the black one). 
For the $\theta$-rotations, Fig.~\ref{f4},  
the difference relates to a slightly
modified contact geometry. 

(d) The contact difference between ``on-'' and ``off-'' states
also manifests itself in the zero bias conductance; according to our
calculations $G_{\rm off}{=}0.0016$ and $G_{\rm on}{=}0.00175$, so 
$G_{\rm off}\apprle G_{\rm on}$. 
Together with (c), the inequality implies that there 
must be an intersection 
of the IVs at a voltage $V_{\rm c}$, which is a detail 
observed in our data, Fig.~\ref{f6}, upper inset. 

\section{$\theta$-invariance of $\Phi$-junctions and IV}
For a more general class of molecular $\Phi-$junctions, 
which do not share the (approximate) symmetries of
the extended molecule, Fig.~\ref{f4}, 
${\cal I}-$invariance can be strongly broken if polarization
effects become large. In this case the hysteretic features of the IV
will be even {\it more} pronounced. Indeed, there is an intrinsic residual 
symmetry of any $\Phi$-junction, 
\begin{equation}
 I_{\rm on}(V_{\rm bias}) =-I_{\rm off}(-V_{\rm bias}),
\label{e2}
\end{equation}  
which in reality will be violated but only due to contacts. 
If the impact of contacts is not too strong, then the approximate validity 
of Eq. (\ref{e2}) ensures, that the junction's
IV-curves closely resemble the switching characteristic of Fig.~\ref{f6}
(main plot) in the sense that there is a well defined 
upper (``on'')  and a lower (``off'') curve. 

\section{Discussion}
\begin{figure}[t]
\begin{center}
\includegraphics[width=0.75\textwidth]{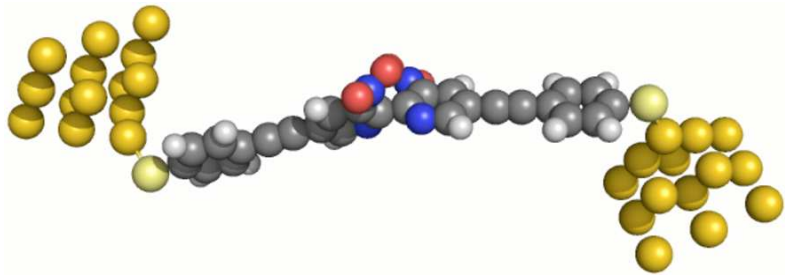}
\end{center}
\caption{Extended molecule used for the DFT
      calculations. The NO$_2$ groups (red/blue) introduce a dipole
      moment perpendicular to the wire direction. In addition, their
      steric interaction drives the wire out of the planar conformation.}
\label{f2}
\end{figure}

\begin{figure}[t]
\begin{center}
\includegraphics[width=0.75\textwidth]{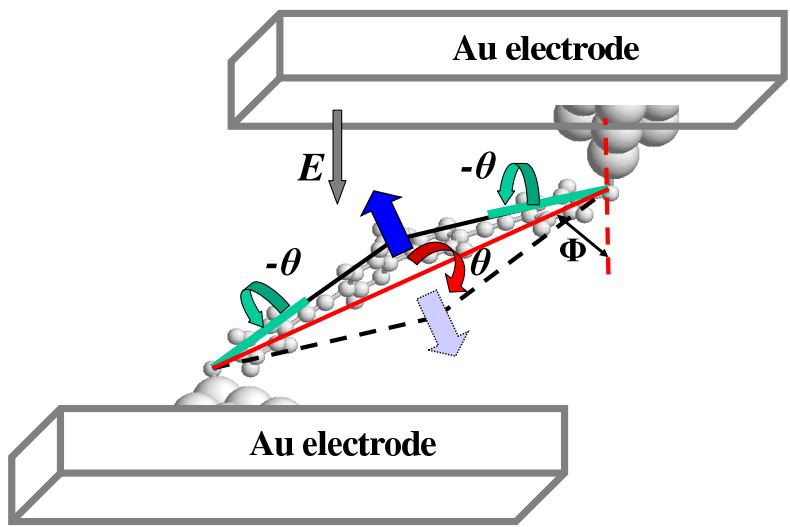}
\end{center}
\caption{Schematics indicating the two axis rotation,
    which connects the two bistable states realizing the molecular
    switch. Blue arrow indicates the NO$_2$ group, grey lines the
    carbon backbone, cf.\ Fig.~\ref{f2}. First step is rotation of the
    whole molecule by $\theta$ about the red axis connecting two
    $S$-atoms. Second step is (back-)rotation of the connecting phenyl ring
    only about the green axis with angle $-\theta$. Back rotation
    partly eliminates the energy cost associated with contact modification.
    Grey arrow indicates the direction of an external electric field.}
\label{f4}
\end{figure}

\begin{figure}[t]
\begin{center}
\includegraphics[width=0.75\textwidth]{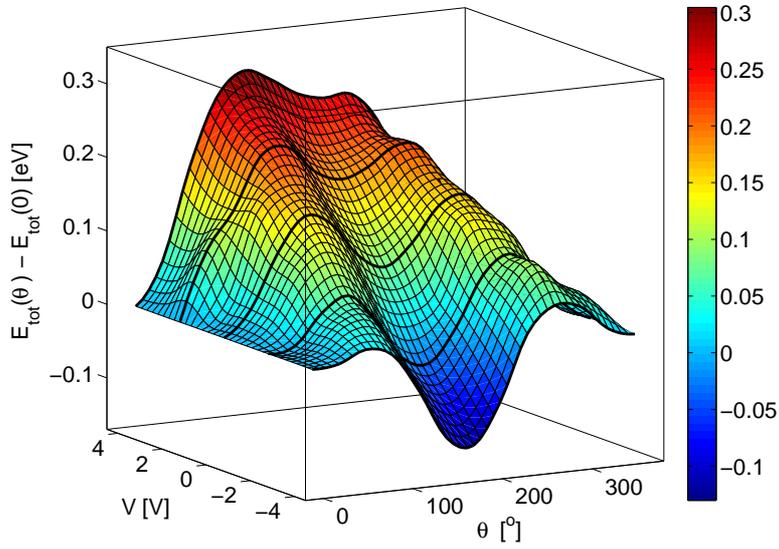}
\end{center}
\caption{Dependency of the total ground state (DFT) energy,
$\Delta E_{\rm tot}(\theta)$, of the extended molecule
(Au$_{14}$-clusters) on the rotation  angle $\theta$
and an inter-electrode electric potential drop, $V$,
with homogeneous gradient ${\bf E}=-\nabla V$, see Fig.~\ref{f4}.
(The simulation angle is $\Phi{=}70^{\rm o}$; full geometry
relaxation has been done for $V = 0$~V at $\theta = 0$.)
}
\label{f5}
\end{figure}

The theoretical IV, Fig.~\ref{f6} shares all essential 
qualitative features with the experimental result \cite{loertscher06}. 
For three reasons, a detailed quantitative comparison 
of absolute values for currents and the characteristic 
voltages, $V_{\rm c}$ and $V_{\rm on}$, appears quite difficult. 
(1) Our theoretical analysis implies that the angle 
$\Phi$ determines the switching voltage, $V_{\rm on}$,  
since it controls the effective force 
($\sim \sin \Phi$) that acts on the intrinsic BPDN-dipole 
moment facilitating the molecular rotation. 
Since details of the atomic conformation of the 
molecular junction realized in experiments are not known, 
$\Phi$ remains largely unspecified. In fact, our atomistic modeling of
the contact, Fig.~\ref{f2}, can be only a crude caricature of 
experimental reality and therefore parameters like $\Phi$ are 
not necessarily very well defined; in addition 
their statistics may fluctuate between different 
experiments \cite{blum05,loertscher06,keane06}. 
(2) Another uncertainty relates to the fact that in experiments the
molecule may not be in a fully relaxed conformation. Our preliminary
calculations show that depending on whether or not molecular
geometries are relaxed, the $E_{\rm tot}(\theta)$-traces depicted in 
Fig.~\ref{f5} vary, if only in a quantitative way. Moreover, our 
switching path, Fig.~\ref{f2}, is a particularly simple, one parameter
species and there is no reason to expect that it is already
optimal. Rather, paths realized in experiments will most likely be
somewhat different, i.e.\ some (smooth) deviations of it. 
(3) It is a well known fact, that due to the local approximations in the 
exchange correlation functionals (XC), DFT calculations do not usually
give precise quantitative results for conductances \cite{evers_arnold} 
or molecular polarizabilities \cite{faassen02}. For this reason, 
our theoretical results have an intrinsic quantitative uncertainty.  

We believe, that in view of points (1-3) it is not surprising that 
our estimate for the switching value 
$V_{\rm on}{\approx}2-4$~V 
does not very precisely reproduce the experimental one 
$\sim 1$~V \cite{blum05,loertscher06}. 

\section{Predictions}
\begin{figure}[t]
\begin{center}
\includegraphics[width=0.35\textwidth,angle=270]{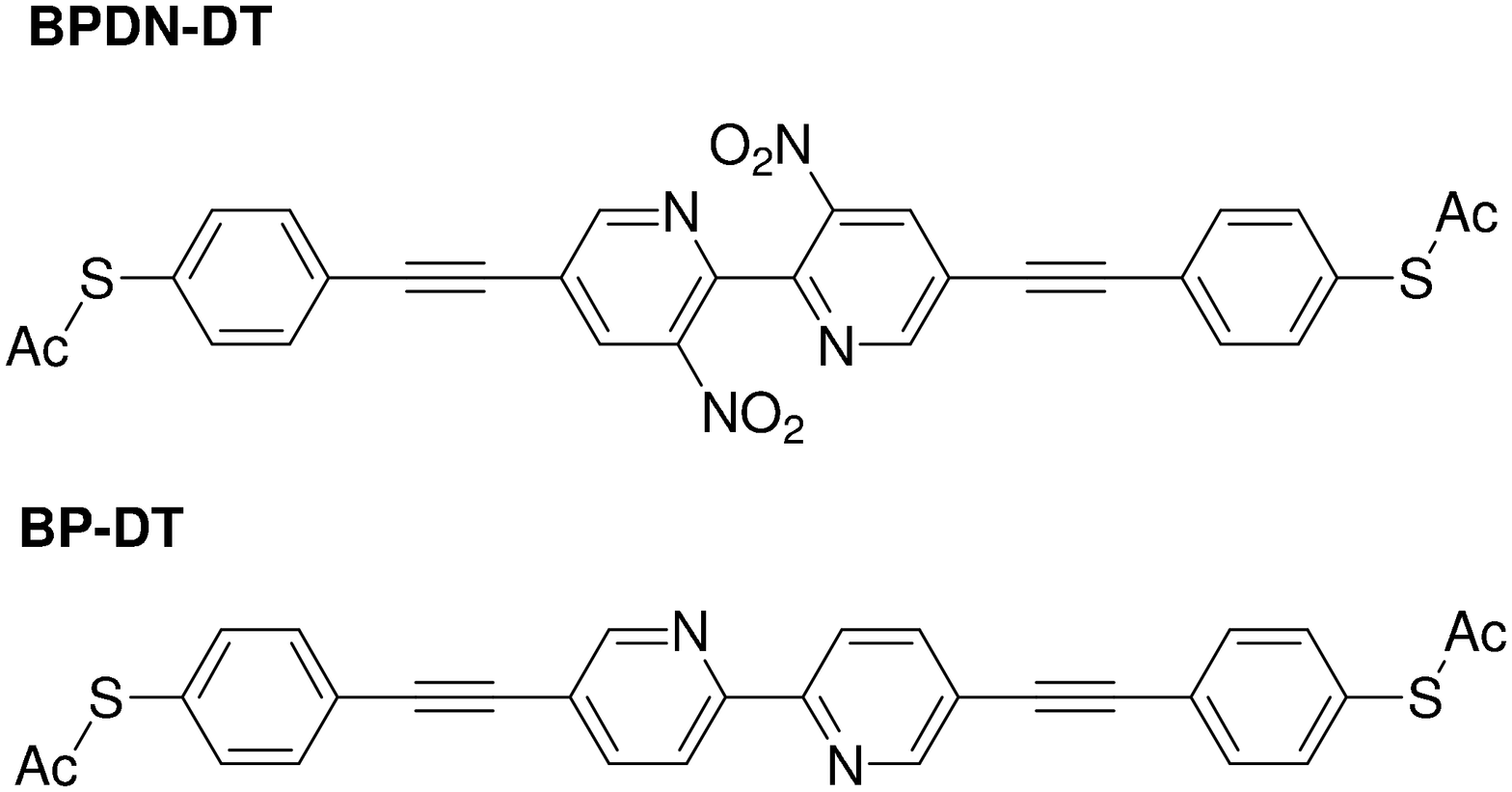}
\end{center}
\caption{Upper panel: The bipyridyl-dinitro
oligophenylene-ethynylene dithiol (BPDN-DT) molecule that shows
switching in single molecule transport experiments
\cite{he06,keane06,loertscher06}.
The protection groups, Ac, are released when
the molecule forms a chemical bond with the Au-surface.
Lower panel: a reference molecule that does not switch \cite{keane06,loertscher06}.}
\label{f1}
\end{figure}

\begin{figure}[t]
\begin{center}
\includegraphics[width=1\textwidth]{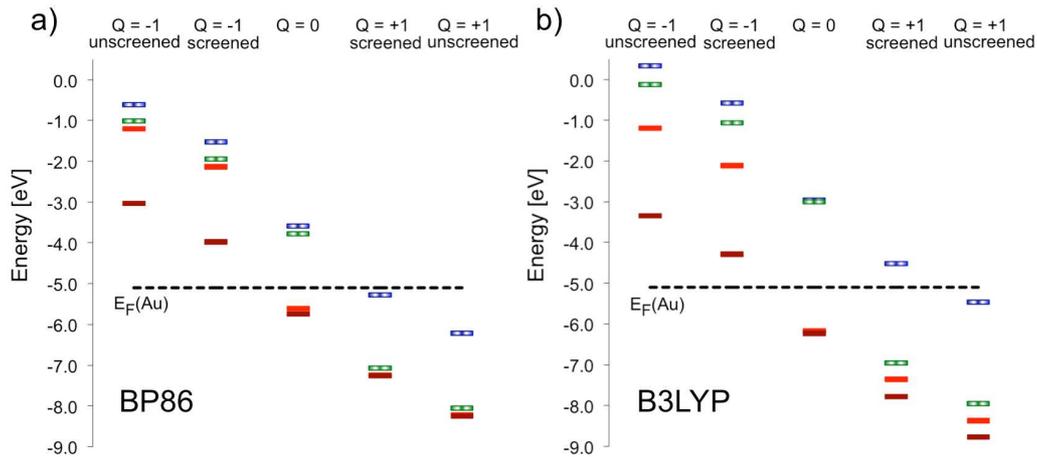}
\end{center}
\caption{\label{t1}
Energies of the highest occupied (HOMO, solid, red) and the lowest unoccupied
(LUMO, dashed, green) molecular levels,
for the molecular wire depicted in Fig.~\ref{f1}
(with the replacement Ac$\to$H). Also neighboring
orbitals, HOMO-1 (solid, brown) and LUMO+1
(dashed, blue), are shown.
For the molecule with excess charge, $Q{=}\pm1$, orbitals have been
obtained with (``screened'') and without (``unscreened'')
image charges; this models screening in the vicinity of metallic (Au)
electrodes. Fermi energy of Au fcc-clusters are
$E_{\rm Fermi}{\approx}-5.05$eV,
with BP-functional (cf.\ $E_{\rm Fermi}{=}-5.65,-5.55,-5.42$~eV for
Au(111), Au(100), Au(110) with local density approximation (LDA)~\cite{umeno07}, 
respectively). The plot on the left (a) shows results obtained within
the BP86 \cite{BP86a,BP86b} exchange correlation functonal, while
the plot of the right (b) shows corresponding results for the B3LYP
\cite{B3LYP} functional. No qualitative difference in level positions
is observed indicating that self-interaction errors do not interfere.
}
\label{f3}
\end{figure}

We close with offering two qualitative predictions 
that follow from our theoretical analysis: 
(p1) The ``easy axis'' rotational degree of freedom of BPDN around the carbon 
triple bonds, Fig.~\ref{f1}, is crucial for $\theta$-rotations. 
If easy rotation is blocked e.~g. by molecular design, then
the rotational barriers become much higher, switching is more 
difficult and hysteresis should be suppressed. 
(p2) The switching voltage, $V_{\rm on}$ can be modified by dressing
the BP-unit, Fig.~\ref{f1}, with other redox groups, e.g.\ 
carboxy-groups, COOH, instead of the nitro-groups, NO$_2$. 
We expect, that the switching voltage decreases when the intrinsic 
dipole moment of the BP-complex becomes larger. 
(Here, we have assumed that the change in molecular conformation,
which accompanies the exchange of the redox groups, is less important.)

Summarizing, we have identified a two-axis rotation that takes 
a Tour-type, BPDN-DT, molecular junction 
from its ground-state conformation (``off'') into
a metastable one (``on''). The energy barrier for this process is larger than
temperature, but can be controlled by applying a bias voltage. 
At sufficiently large biases, the molecule undergoes the transition from 
``off''-state to ``on'' and reverse, so that it can realize a molecular
switch. Our theoretical analysis may explain the pronounced switching
behavior observed in recent experiments 
\cite{blum05,keane06,loertscher06}.  

\section*{Acknowledgments}  
We gratefully acknowledge helpful discussions with 
K.\ Fink, H.\ Fliegl, M.\ Mayor, 
M.\ Ruben, F.\ Schramm and Th.\ Wandlowski.
Our particular gratitude belongs to our colleagues at IBM Zurich, 
E.\ L\"ortscher and H.\ Riel, for inspiring discussions and  
for sharing generously unpublished material. 
This work was supported by the Center of
Functional Nanostructures at Karlsruhe University. 



\clearpage

\clearpage

\clearpage

\clearpage

\clearpage

\clearpage

\clearpage
\section*{Table of Contents}

Recent experiments have identified a specific molecular wire 
(bipyridyl-dinitro oligophenylene-ethynylene dithiol), BPDN-DT that can 
be operated as a molecular memory element. Our theoretical work
explains the mechanism by which the memory is conserved. At its heart 
it is a two-axis rotation of the molecule’s functional unit (see Figure). 
The theoretical current-voltage characteristics reproduces the experimental 
observations.

{\bf ToC keyword:} Molecular switches\\

{\bf Velimir Meded, Alexei Bagrets, Andreas Arnold, and Ferdinand Evers$^*$}\\

{\bf Molecular switch controlled by pulsed bias voltages}

\begin{figure}[h]
\begin{center}
\includegraphics[width=55mm]{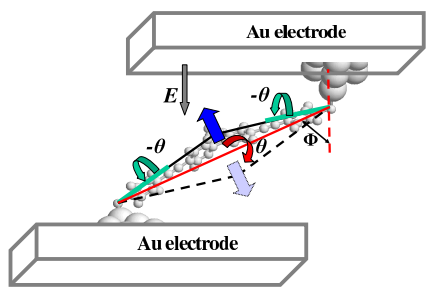}
\end{center}
\caption{Table of Contents Graphic (Figure 4., in color) }
\label{ToC Graphic}
\end{figure}

\end{document}